%% file: paper.tex
\documentclass[11pt,a4paper]{article}
\pdfoutput=1
\usepackage{jheppub}
\usepackage{graphicx} 
\usepackage{epstopdf} 
\usepackage{dcolumn} 
\usepackage{xcolor} 
\usepackage{amsmath,amssymb} 
\usepackage{hyperref} 
\usepackage[utf8]{inputenc} 
\usepackage[english]{babel} 
\usepackage{blindtext} 
\usepackage{ifthen} 
\usepackage{orcidlink} 
\usepackage{multirow}
\usepackage{float}
\graphicspath{{figures/}} 

\newboolean{articletitles}
\setboolean{articletitles}{true} 

\input{belle2-symbols}

\def\eetautau     {\ensuremath{e^+e^-\to\tau^+\tau^-}\xspace}

\def\eeqqbar     {\ensuremath{e^+e^-\to q \bar q}\xspace}
\def\taumu    {\ensuremath{\tau^-\to \mu^- \mu^+ \mu^-}\xspace}
\def\Mmu {\ensuremath{M_{3\mu}}\xspace}
\def\dE {\ensuremath{\Delta E_{3\mu}}\xspace}
\def\visE  {\ensuremath{{E_{\rm vis}^*}}}

\begin{document}

\vspace*{-3\baselineskip}
\resizebox{!}{2cm}{\includegraphics{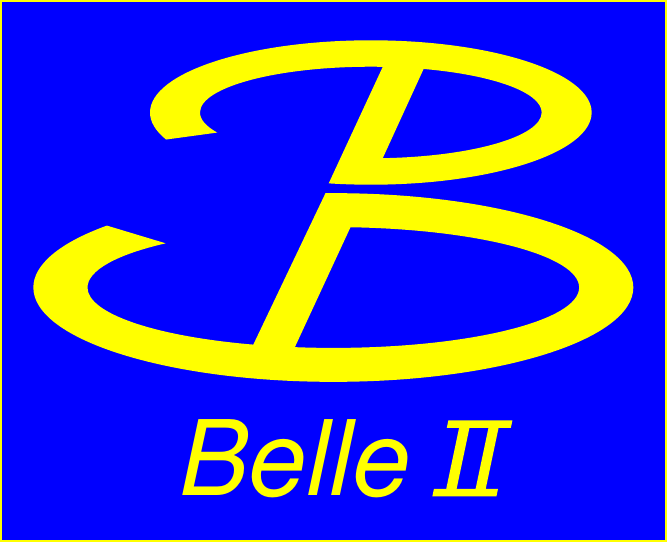}}
\begin{flushright}
Belle II Preprint 2024-012 \\
KEK Preprint 2024-6
\end{flushright}
\title{Search for lepton-flavor-violating $\tau^- \to \mu^-\mu^+\mu^-$ decays at  Belle II}

\collaboration{The Belle II Collaboration}
  \author{I.~Adachi\,\orcidlink{0000-0003-2287-0173},} 
  \author{L.~Aggarwal\,\orcidlink{0000-0002-0909-7537},} 
  \author{H.~Aihara\,\orcidlink{0000-0002-1907-5964},} 
  \author{N.~Akopov\,\orcidlink{0000-0002-4425-2096},} 
  \author{A.~Aloisio\,\orcidlink{0000-0002-3883-6693},} 
  \author{N.~Althubiti\,\orcidlink{0000-0003-1513-0409},} 
  \author{N.~Anh~Ky\,\orcidlink{0000-0003-0471-197X},} 
  \author{D.~M.~Asner\,\orcidlink{0000-0002-1586-5790},} 
  \author{H.~Atmacan\,\orcidlink{0000-0003-2435-501X},} 
  \author{V.~Aushev\,\orcidlink{0000-0002-8588-5308},} 
% ...

\emailAdd{coll-publications@belle2.org}
\abstract{We present the result of a search for the charged-lepton-flavor violating decay \taumu using a 424\invfb sample of data recorded by the Belle II experiment at the SuperKEKB \epem collider.
The selection of $\epem\to\tau^+\tau^-$ events is based on an inclusive reconstruction of the non-signal tau decay, and on a boosted decision tree to suppress background.
We observe one signal candidate, which is compatible with the expectation from background processes. We set a $90\%$ confidence level upper limit of $1.9 \times  10^{-8}$ on the branching fraction of the \taumu decay, which is the most stringent bound to date.}

\maketitle
\flushbottom

\input{01-intro}
\input{02-belleII}

\input{03-selection}
\input{04-sys}
\input{05-UL}
\input{06-summ}

\input{acknowledgements-b2}

\bibliographystyle{JHEP}
\bibliography{references_jhep}
\newpage
\appendix
\section*{Additional Material}
\begin{figure}[H]
    \centering
  
    \includegraphics[page=14,width=0.8\columnwidth]{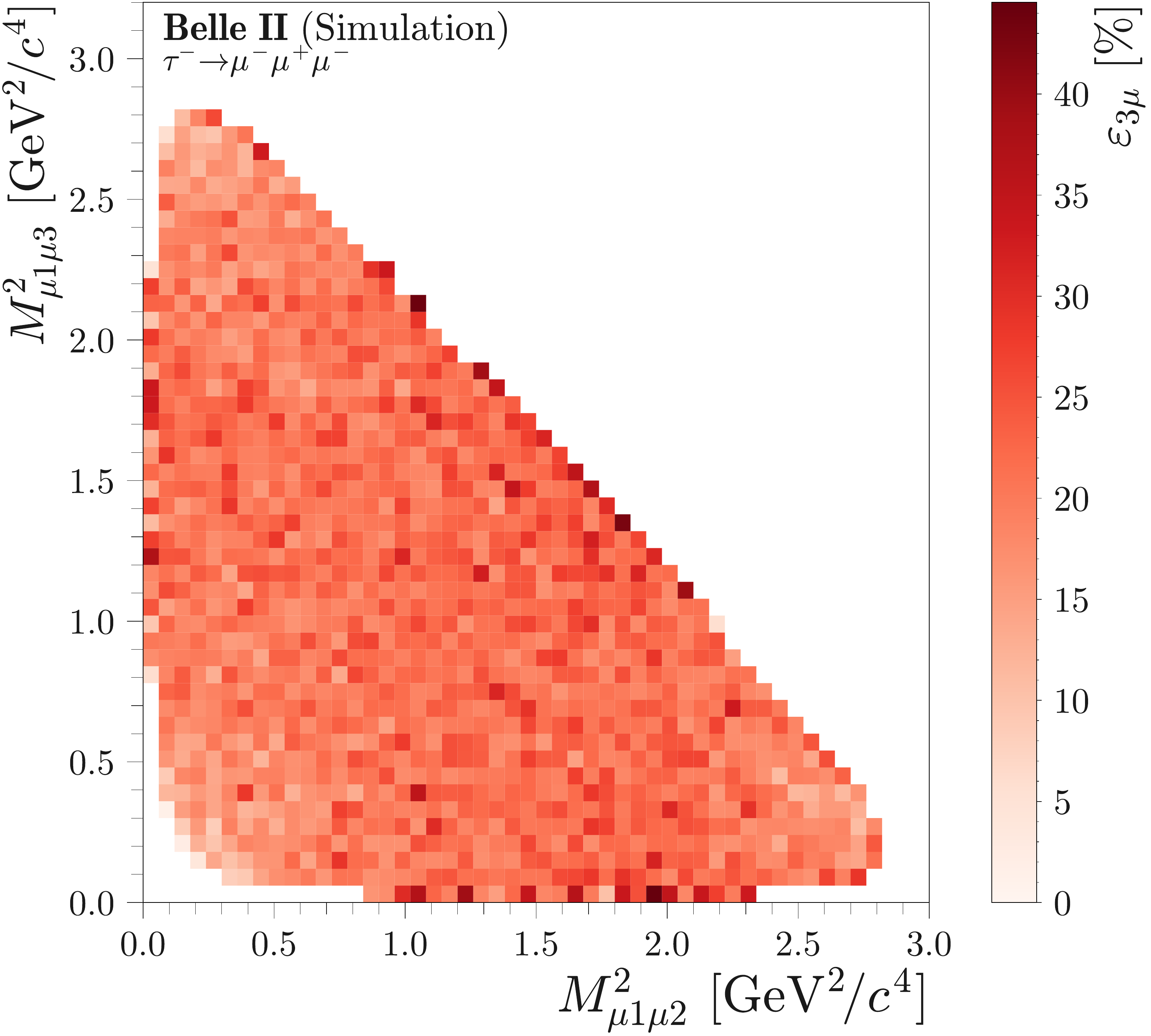}
    \caption{Signal efficiency distribution in the two-dimensional plane defined by the mass squared of the opposite charge muons, for candidates passing the full selection.}
    \label{fig: Dalitz plots}
\end{figure}

The signal efficiency as function of the two-dimensional plane defined by the mass squared of the opposite charge muons is provided in  Fig.~\ref{fig: Dalitz plots}.

\begin{figure}
    \centering
    \includegraphics[width=0.49\textwidth,page=9]{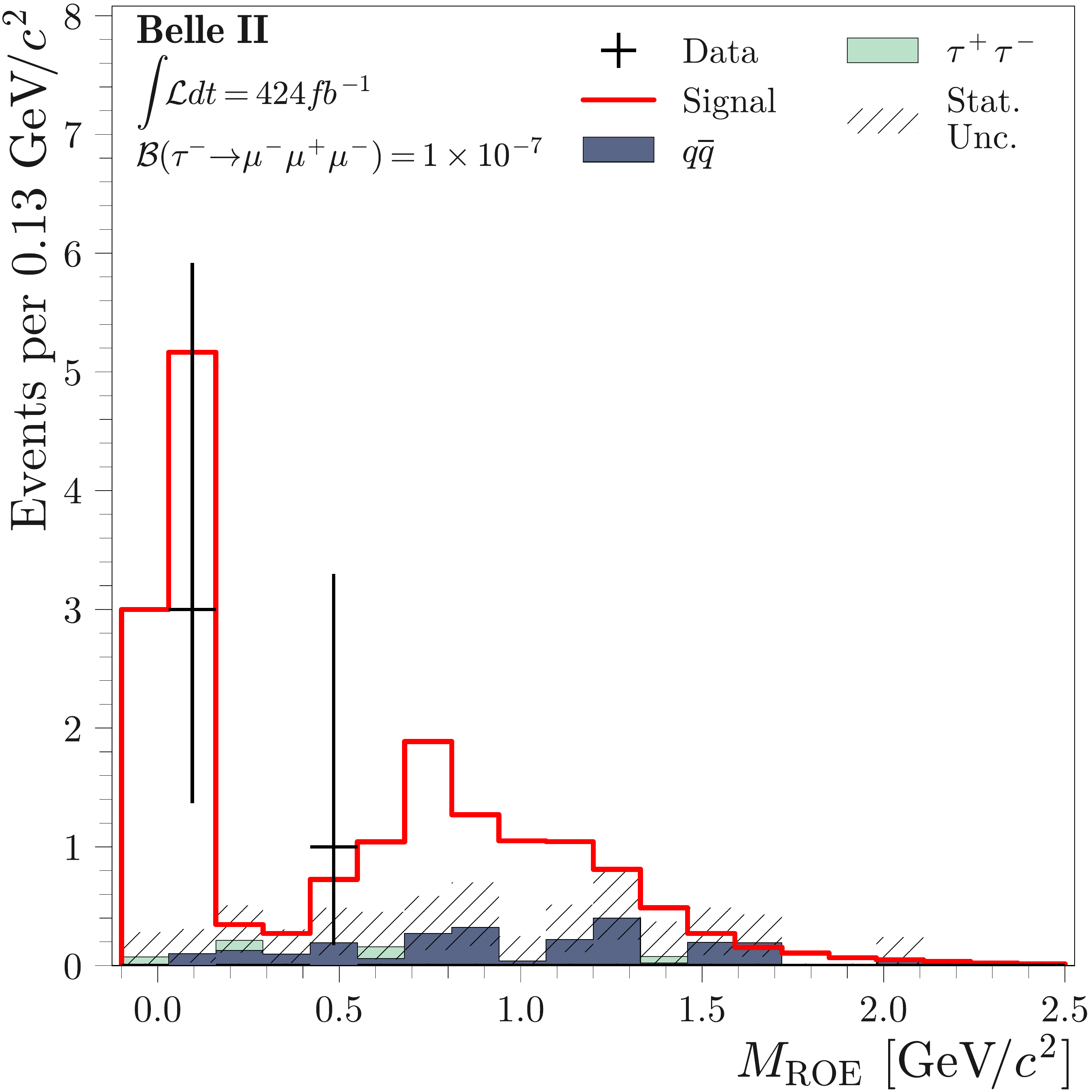}
    \includegraphics[width=0.49\textwidth,page=10]{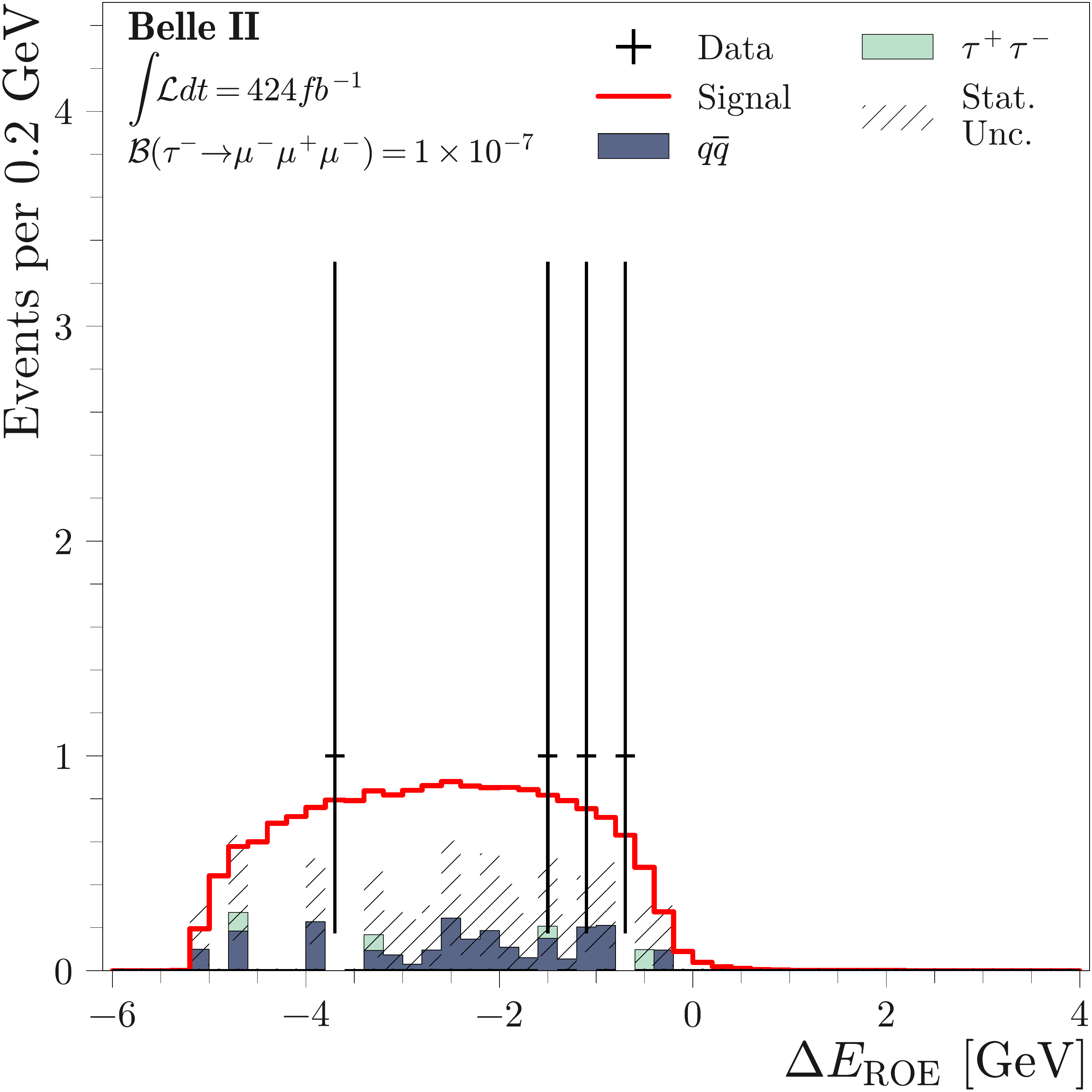}
    \includegraphics[width=0.49\textwidth,page=11]{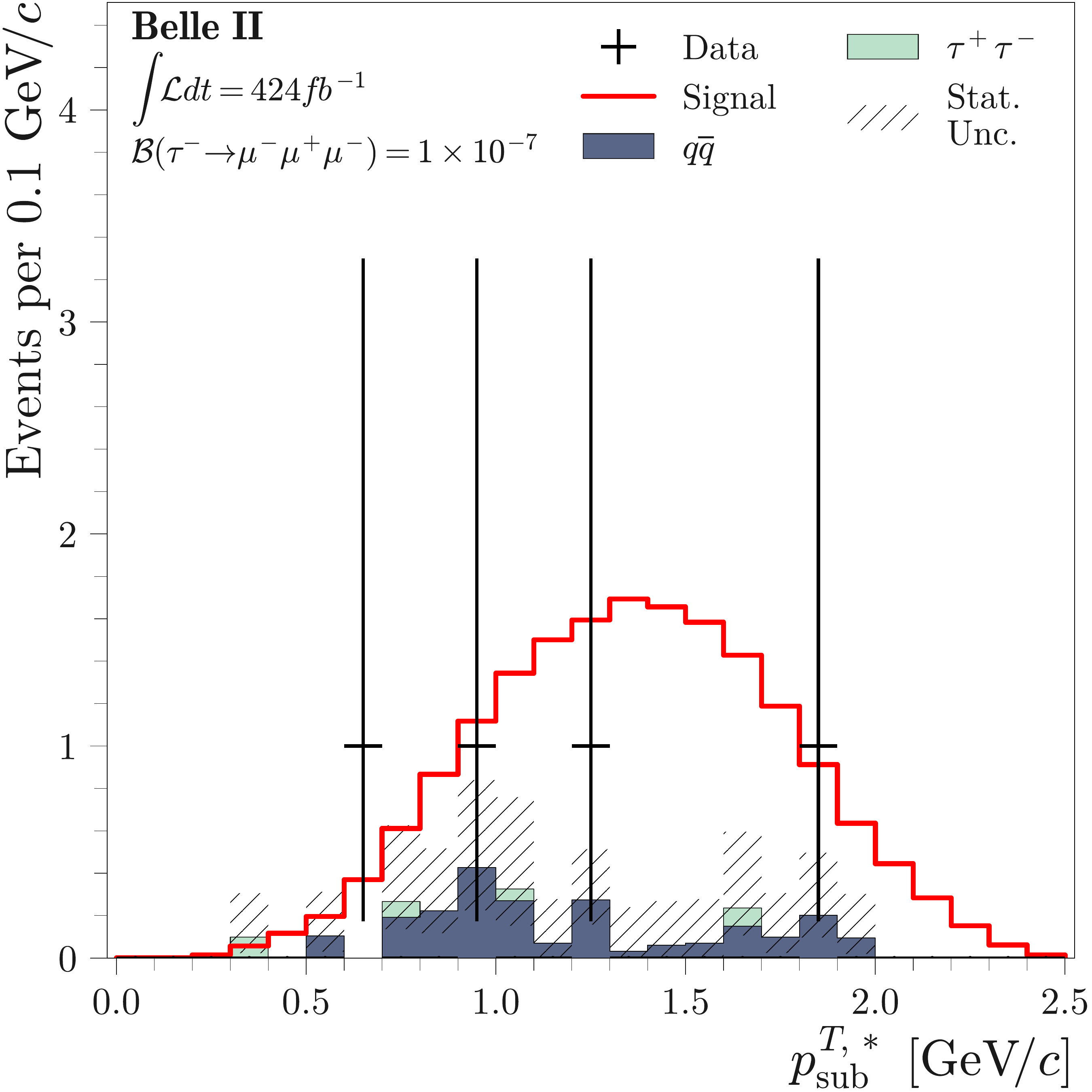}
    \includegraphics[width=0.49\textwidth,page=12]{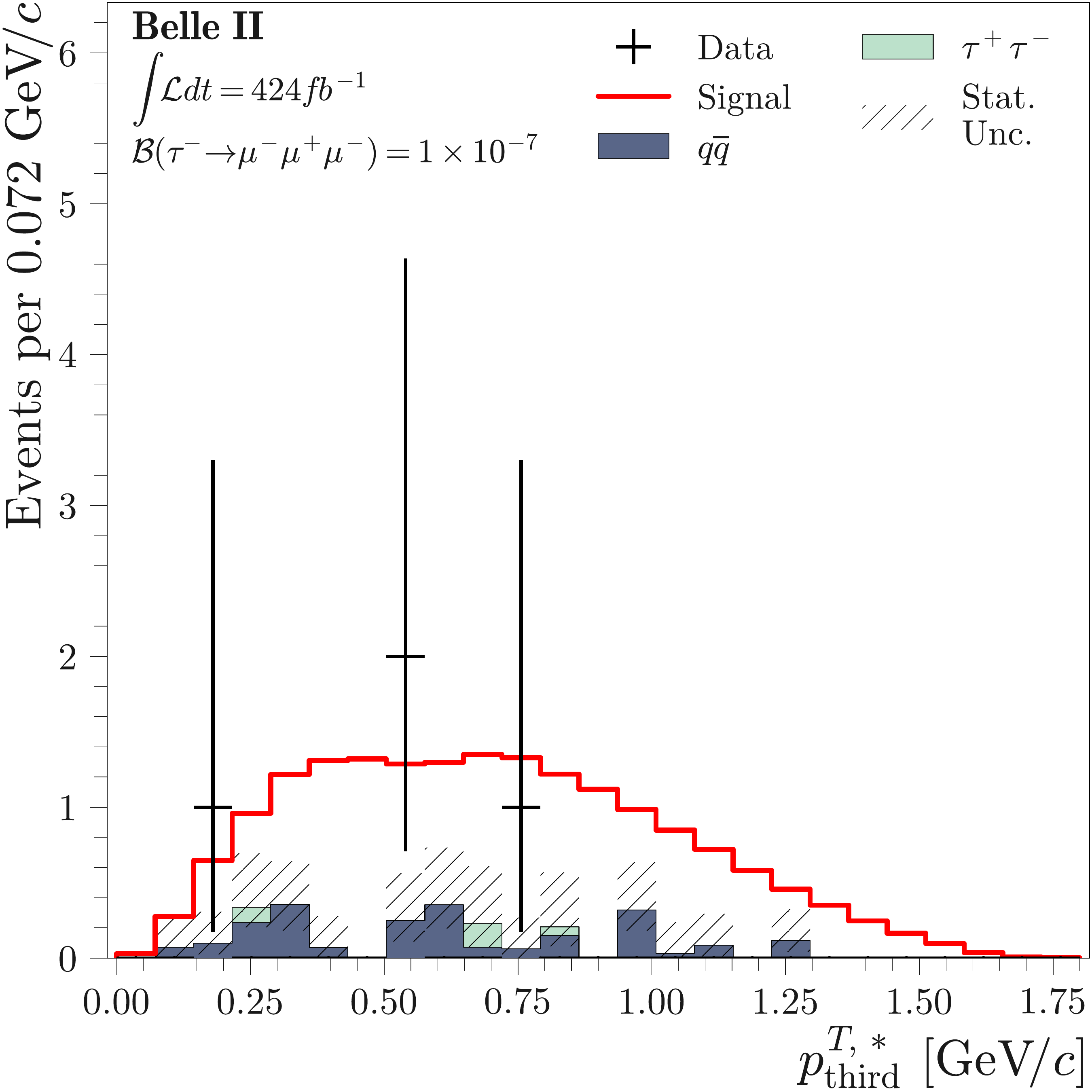}
    \caption{Comparison between data (black points with error bars) and simulation for the most discriminating variables used in the BDT: mass of the rest of the event (upper-left), difference of energy between the rest of event and the beam (upper-right), transverse momentum of the second highest momentum muon (bottom-left) and transverse momentum of the lowest momentum muon (bottom-right). Events in the sideband and signal regions are used. The various simulated background processes are shown as a stack of color-filled histograms, while the signal is shown as a red histogram with an arbitrary scale. The statistical uncertainties are displayed as the grey-hatched areas.}
    \label{fig: Inclusive DataMC comparison for best BDT variables}
\end{figure}

Comparison between data and simulation for the most discriminating variables used in the BDT are shown in Fig.~\ref{fig: Inclusive DataMC comparison for best BDT variables}.
\end{document}

%% file: 01-intro.tex
\section{Introduction}
\label{sec:intro}

In the standard model (SM) with massless neutrino hypotheses, the charged-lepton flavor is accidentally conserved.
However, this symmetry is broken at loop-level when taking into account neutrino mixing, which implies the existence of charged-lepton-flavor violation, and thus the existence
of processes such as $\mu\to e$, $\tau\to e$ and  $\tau\to \mu$ conversions.
In the simplest SM extension that allows for massive neutrinos, all charged-lepton-flavor-violating (LFV) rates are proportional to the square of neutrino masses. This results in predicted decay rates of $10^{-50}$~\cite{ref:SM_cLFV, ref:tau_lfv, Blackstone:2019njl}, well below the sensitivities of any experiment. The observation of  LFV decays would, therefore, provide indisputable evidence of physics beyond the SM.

Over the past four decades, the CLEO experiment at CESR, and the first generation $B$-factory experiments BaBar at SLAC and Belle at KEK, have searched for LFV in $\tau$-lepton decays. In total, 52 LFV $\tau$ decays with neutrinoless two-body or three-body final states have been investigated.
Among these,  $\tau^-\to\ell^-\ell^{+\prime}\ell^{-\prime\prime}$ decays,\footnote{Charge conjugation is implied throughout this paper.} where $\ell^{(\prime,\prime\prime)}=e,\mu$, and in particular \taumu{}, have garnered significant attention in recent years. This is due to the potential enhancement of the branching fraction up to a value of $10^{-8}$ in scenarios beyond the SM~\cite{Abada:2021zcm, Raidal:2008jk, Teixeira:2016ecr, Bhattacharya:2016mcc, Feruglio:2017rjo, Greljo:2015mma}. 
The most stringent upper limit for this decay was set by the Belle collaboration at the level of $2.1\times10^{-8}$ at the 90\% confidence level (C.L.) using electron-positron data corresponding to an integrated luminosity of  782\invfb~\cite{Hayasaka:2010np}.
The upper limits on the \taumu branching fraction set by the CLEO~\cite{CLEO:1997aqy} and BaBar~\cite{BaBar:2010axs} collaborations are 1.9$\times10^{-6}$ and $3.3\times10^{-8}$,  using 4.79\invfb and  468\invfb of data, respectively. 
In addition, experiments at the Large Hadron Collider have also contributed to this search. The LHCb~\cite{LHCb:2014kws}, ATLAS~\cite{ATLAS:2016jts}, and CMS~\cite{CMS:2020kwy} collaborations reported upper limits at the 90\% C.L. of $4.6\times10^{-8}$, $3.8\times10^{-7}$ and $8.0\times10^{-8}$ using proton-proton collision data corresponding to integrated luminosities of 3\invfb, 20.3\invfb, and 33.2\invfb respectively. 
Recently, CMS  presented a new result incorporating 2017 and 2018 data, amounting to a total integrated luminosity of 131\invfb, placing an upper limit at 2.9  $\times10^{-8}$ at 90\% C.L~\cite{CMS:2023iqy}.

We report the results of a search for the LFV \taumu{} decay using an untagged selection, in which,
in contrast to the previous searches, only the signal tau decay mode is explicitly reconstructed.
We use a sample of 389 million $e^+e^-\to\tau^+\tau^-$ events recorded with the \belletwo{} detector~\cite{Abe:2010gxa} at the asymmetric-energy \epem{} SuperKEKB collider~\cite{Akai:2018mbz}.
The data, collected between 2019 and 2022, corresponds to an integrated luminosity of 424\invfb.
Section~\ref{sec:BelleII} gives an overview of the Belle II detector and the data samples used. Section~\ref{sec:selection} presents the overall strategy for this analysis and the details of the candidate reconstruction and selection.
Section~\ref{sec:systematics} discusses the systematic uncertainties, and Section~\ref{sec:results} presents the branching fraction measurement and limit computation. Finally, a summary is given in Section~\ref{sec:summ}.

%% file: 02-belleII.tex
\section{The Belle II detector, simulation and data samples}
\label{sec:BelleII}
The Belle II detector consists of several subdetectors arranged in a cylindrical structure around the $e^+e^-$ interaction point~\cite{Abe:2010gxa}. 
Charged-particle trajectories (tracks) are reconstructed using a two-layer silicon-pixel detector, surrounded by a four-layer double-sided silicon-strip detector and a central drift chamber (CDC). Only 15\% of the second pixel layer was installed when the data were collected. 
Outside the CDC, which also provides $d{E}$/$d{x}$ energy-loss measurements, particle identification using Cherenkov radiation is provided by a time-of-propagation (TOP) detector and an aerogel ring-imaging Cherenkov (ARICH) detector which cover the barrel and forward endcap regions, respectively. 
An electromagnetic calorimeter (ECL), divided into the forward endcap, barrel, and backward endcap regions, fills the remaining volume inside a 1.5 T superconducting solenoid and is used to reconstruct photons and electrons.  
A $K_L^0$ and muon detection system (KLM) is installed in the iron flux return of the solenoid.  
The $z$ axis of the laboratory frame is defined as the detector solenoid axis, with the positive direction along the electron beam. The polar angle $\theta$ and the transverse plane are defined relative to this axis.

The search presented here is based on \epem{} collisions at the center of mass energy of 10.58 GeV corresponding to the mass of the \FourS (362\invfb), 60~\mev below it (42\invfb), and around 10.75 GeV  (19\invfb). The corresponding cross sections for \eetautau{} production are $\sigma_{\tau \tau}$=0.919~\nb at the \FourS energy,   0.929~\nb for off-resonance data, and   0.891~\nb around the \FiveS energy,
leading to a data sample of 389 million $\tau$-pairs~\cite{Banerjee:2007is}. 

Monte-Carlo simulated events are used to optimize the selection and background rejection and to measure the signal efficiency. To study the signal process, we use 10 million  \eetautau events, in which one $\tau$ decays to three muons following a phase space model and the other to SM-allowed decays.
The potential background processes studied using simulation include $\epem \to \qqbar$ events, where $q$ indicates a $u$, $d$, $c$, or $s$ quark;  $\epem \to b\Bar{b}$ events; $\epem\to \ell^+ \ell^- (\gamma)$, where $\ell=e,\mu$; $\epem \to e^+e^-h^+h^-$ events, where $h$ indicates a pion, kaon, or proton; and four-lepton processes: $\epem \to e^+e^-e^+e^-,\mu^+\mu^-\mu^+\mu^-,\mu^+\mu^-e^+e^-,e^+e^-\tau^+\tau^-, \mu^+\mu^-\tau^+\tau^-$. The \eetautau{} process is generated using the KKMC generator~\cite{Jadach:1999vf}. The $\tau$ decays are simulated by the TAUOLA generator~\cite{Jadach:1990mz} and their final state radiated photons by PHOTOS~\cite{Barberio:1990ms}. 
We use KKMC to simulate $\mu^+\mu^-(\gamma)$ and \qqbar production; PYTHIA~\cite{Sjostrand:2014zea} for the fragmentation of the \qqbar pair; PYTHIA interfaced with EvtGen~\cite{Lange:2001uf} for the production and decay of $\epem \to b\Bar{b}$ events; BabaYaga@NLO~\cite{Balossini:2006wc, Balossini:2008xr, CarloniCalame:2003yt, CarloniCalame:2001ny,CarloniCalame:2000pz} for $\epem \to \epem (\gamma)$ events; and AAFH~\cite{BERENDS1985421,BERENDS1985441,BERENDS1986285} and TREPS~\cite{Uehara:1996bgt} for the production of non-radiative four-leptons and $\epem h^+h^-$ final states. 
The size of the simulated samples for  \eetautau{}
and $\epem \to \qqbar$ events is equivalent to an integrated luminosity of 8 ab$^{-1}$, while it ranges between 100\invfb and 2 ab$^{-1}$ for the other processes.
The \belletwo{} analysis software~\cite{Kuhr:2018lps, basf2-zenodo} uses the GEANT4~\cite{Agostinelli:2002hh} package to simulate the response of the detector to the passage of the particles and also provides a simulation of the triggers.

The online event selection (hardware trigger) is based on the energy deposits (clusters) and their topologies in the ECL,  or on independent trigger selection based on the number of charged particles reconstructed in the CDC.
Most of the events are selected requiring a ECL total energy larger than 1 GeV and a topology incompatible with Bhabha events.

%% file: 03-selection.tex
\section{Event reconstruction and background rejection}
\label{sec:selection}

\subsection{Overall strategy}

In the \epem center-of-mass (c.m.)\ frame, the $\tau$ leptons are produced in opposite directions, with the decay products of one $\tau$ isolated from those of the other $\tau$ and contained in opposite hemispheres. 
The boundary between the hemispheres is experimentally defined by the plane perpendicular to the vector $\mathbf{\hat{n}}_T$ that maximizes the thrust value ($T$):\begin{equation}
\label{eq:thrust}
	T = \max_{\mathbf{\hat{n}}_T} \left(\dfrac{\sum_{i} \left|\mathbf{p^*}_i \cdot \mathbf{\hat{n}}_T\right|}{\sum_{i} \left|\mathbf{p^*}_i\right|} \right),
\end{equation}
where $\mathbf{p}_i^*$ is the momentum of final state particle $i$ in the \epem{} c.m.\ frame~\cite{Brandt:1964sa,Farhi:1977sg}, including both charged and neutral particles. Here, and throughout the paper, quantities in the $\epem$ c.m.\ frame are indicated by an asterisk.

We define the signal hemisphere as the one that contains the \taumu{} decay candidate, reconstructed by combining three charged particles. The trajectories of those particles are required to be displaced from the average interaction point by less than 3~\cm along the $z$ axis and less than 1~\cm in the transverse plane. 
Identification of muons of momentum greater than 0.7~\gevc  relies mostly on their penetration depth in the KLM whereas for lower momenta information from the CDC and ECL dominate. Muons are identified using the discriminator $\mathcal{P}_\mu = {\cal L}_\mu / ({\cal L}_e + {\cal L}_\mu + {\cal L}_\pi + {\cal L}_K + {\cal L}_p + {\cal L}_d)$ where the likelihoods ${\cal L}_i$ for each charged-particle hypothesis $i = e, \mu, \pi, K, \text{proton} (p), \text{deuteron} (d)$ combine particle-identification information from CDC, TOP, ARICH, ECL, and KLM. 
The muon identification efficiency and misidentification rates are measured from independent samples as function of the particle momentum: di-photon events are used for the efficiency measurement below 1.5 GeV/c while $J/\Psi$  and di-muon events
are used for higher momenta. As an example, the muon identification efficiency for the requirement $\mathcal{P}_\mu>0.5 (0.95)$ is 92(89)\% with a pion misidentification probability of 5(3)\%, for particle momenta between 1.0 and 1.5\gevc.

Since the decay \taumu{} is a neutrinoless process, the invariant mass \Mmu of the reconstructed muons should be consistent with the mass of the $\tau$ lepton, except for decays affected by final state radiation (FSR) from the $\tau$ or its decay products. The energy $E_\tau^*$ in the c.m.\ system should be half of the \epem c.m.\ energy $\sqrt{s}/2$ except for corrections from initial state radiation (ISR) from the $e^\pm$ beams and  FSR. Thus, the energy difference $\dE = E_\tau^*-\sqrt{s}/2 $ should be close to zero. These characteristic features are used to define the signal region, which is hidden until the finalization of the selection procedure to avoid any experimenter's bias, and to optimize the selection criteria, which maximize the signal efficiency and suppress the contribution of background events.
The distribution of signal in the (\Mmu, \dE) plane is broadened by detector resolution and radiative effects. The radiation of photons from the initial state leads to a tail at low values of \dE while FSR produces a tail at low value both in \Mmu and \dE.

Previous \B-factory experiments relied on the reconstruction of $\epem\to\tautau$ events where one $\tau$ decays to $\mu^-\mu^+\mu^-$ and the other $\tau$, named the \emph{tag}, decays into a final state with a single charged particle. The background rejection strategy relied on a cut-based selection.
Here, we expand the search by using inclusive tagging to
include a wider range of decays for the tag $\tau$ lepton. We also optimize the signal efficiency and reduce the background by using boosted decision trees (BDT). 
To compare the sensitivity of the inclusive tagging to the traditional one-prong tagging, we also perform a validation using a reconstruction method similar to the one used previously by Belle~\cite{Hayasaka:2010np} and BaBar~\cite{BaBar:2010axs}.

\subsection{Inclusive tagging selection}
\input{032-BDTbased}

%% file: 032-BDTbased.tex
Signal candidate \taumu{} decays are obtained by combining three muons originating from the interaction point with a total charge equal to $\pm 1$, belonging to the same hemisphere and having $\mathcal{P}_{\mu}>0.5$. 
Instead of explicitly reconstructing the tagged $\tau$ lepton, we use the inclusive properties of all other particles in the event. Photons are reconstructed from ECL clusters within the CDC acceptance ($17^\circ  < \theta < 150^\circ$) and not associated with any tracks. Photons used for \piz reconstruction must have an energy deposit of at least 0.1~\gev.
Neutral pions are then identified as photon pairs with invariant masses within $0.115 < M_{\gamma\gamma}< 0.152$~\gevcc, which corresponds to a range of approximately $\pm 2.5\sigma$ around the known $\pi^0$ mass~\cite{ParticleDataGroup:2022pth}, $\sigma$ being the  $M_{\gamma\gamma}$ resolution.
All photons with energies greater than 0.2~\gev, along with photons part of reconstructed $\pi^0$ candidates,  are used to define variables related to the kinematic properties of the event such as the missing momentum, defined as the difference between the momenta of the initial $e^+e^-$ and that of all reconstructed tracks and photons in the event,  its mass and the thrust axis. This high energy threshold for photon candidates aims to reduce photons from beam background.
All tracks and clusters that are not used in the signal reconstruction form the rest of the event (ROE), whose kinematic properties are exploited for further background suppression. In order to suppress beam-related backgrounds, the following criteria must be satisfied for particles to be retained as part of the ROE: photons are required to have energies of at least 0.2~\gev, and tracks must have transverse momenta higher than 0.075~\gevc, and be displaced from the average interaction point by less than 3~\cm along the $z$ axis and less than 1~\cm in the transverse plane. In addition, tracks and photons should lie within the CDC angular acceptance.

We define rectangular regions in the two-dimensional plane (\Mmu, \dE) centered at the expected position of the signal peak ($\Bar{\mu}$), with side lengths proportional to the expected resolutions ($\delta$). For each variable, $\delta$ values are estimated as the widths of a bifurcated Gaussian function fitted to the simulated signal distribution. The expected peak position $\Bar{\mu}$, and the high- and low-mass resolutions $\delta_{high(low)}$, are reported in Table~\ref{tab:sig_resol}.

\begin{table}[htbp]
    \centering
    \caption{Fitted central values and resolutions for \Mmu and \dE.}
    \begin{tabular}{c|ccc}               
    Variable & \newline$\Bar{\mu}$ &  $\delta_{low}$ &  $\delta_{high}$ \\
    \hline
    \Mmu (\mevcc) & 1777.35$\pm$0.07 & 4.80$\pm$0.07 & 4.44$\pm$0.06 \\
    \dE (\mev) & 0.7$\pm$0.3 & 14.9$\pm$0.3 & 10.0$\pm$0.5 \\
    \end{tabular}
    \label{tab:sig_resol}
\end{table}

A $\pm20\,\delta$-wide rectangular region in the (\Mmu, \dE) plane is used for the background-rejection optimization using simulated data. The final yield extraction is performed in a 5 $\delta$ semi-axis wide, asymmetric elliptical signal region (SR). The rotation angle of the axis of the ellipse is obtained from a linear fit to the profile distribution of \Mmu versus \dE.
The sideband region (SB), defined as the area covering $\pm20\,\delta_M$, $\pm10\,\delta_{\Delta E}$ subtracted of the SR, is used for the validation of the background rejection.

\begin{figure}[!ht]
    \centering    
        \includegraphics[width=0.49\textwidth,page=1]{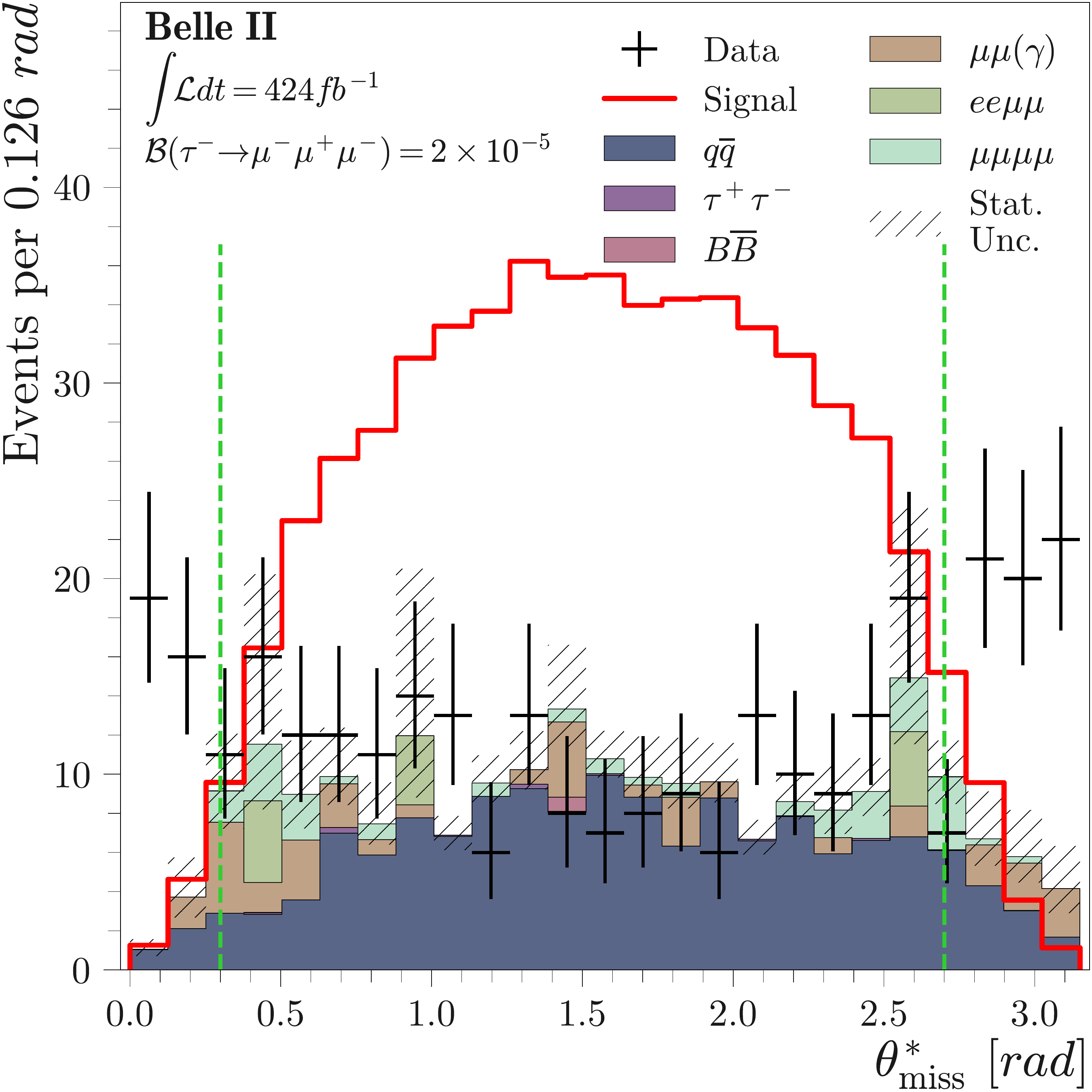}
      \includegraphics[width=0.49\textwidth,page=2]{figures/Plot_forPaper.pdf}
    \caption{Comparison between sideband data (black points with error bars) and simulation after the inclusive tagging reconstruction for the polar angle of missing momentum (left) and thrust (right) distributions.  The various simulated background processes are shown as a stack of color-filled histograms, with statistical uncertainties displayed as hatched areas, while the signal is shown as a red histogram with an arbitrary scale. The vertical dashed green lines indicate the boundaries of the selection criteria.}
    \label{fig:presel}
\end{figure}

The background originates from radiative dilepton and four-lepton final states (low-multiplicity backgrounds) with potential electrons misidentified as muons, incorrectly reconstructed SM $\epem\to\tautau$ events, and continuum hadronization processes from $\epem\to\qqbar$ events, where pions are misidentified as muons. The other simulated processes are found to be negligible.
To suppress events with pions or other particles misidentified as muons, two of the muon candidates are required to have $\mathcal{P}_\mu >0.95$.
Low-multiplicity backgrounds are concentrated at very high thrust values and have a missing momentum pointing at the boundaries of the polar angular acceptance in the c.m.\ frame. In Fig.~\ref{fig:presel}, there are disagreements between data and simulation in these regions, consistent with unsimulated low-multiplicity backgrounds such as four-muon final state processes with initial and final state radiations.
To suppress such backgrounds, we place requirements on the polar angle of the missing momentum, $0.3<\theta^*_{\rm miss}<2.7\rad$, and the thrust,  $0.89<T<0.97$. The overall signal efficiency after those requirements, including event reconstruction and selection, is $30.7\%$, while analysis of the simulated background events predicts about 300 events in the 
$\pm 20\,\delta$ region. They are mostly due to  $\epem\to\qqbar$ processes in which one or two of the final state muons are misidentified.
\subsection{BDT-based background rejection}

To suppress the remaining background events, a boosted decision tree classifier (BDT) is trained using the XGBoost library~\cite{XGBoostPaper}. The BDT uses 32 variables related to three distinct categories.
The first category consists of variables associated with the signal $\tau$, such as the muon energies, polar angles and ordered transverse momenta in the c.m.\ frame; the flight time of the $\tau$ divided by its uncertainty; the cosine of the angle between the $\tau$ momentum and its direction reconstructed from the origin and decay vertices; the angle between the $\tau$ momentum and the closest track in the event (including final-state muons); and the $\chi^2$ of the kinematic fit of the $\tau$ decay chain.
The second category involves variables related to the ROE properties such as its mass, defined as the mass of the four-vector resulting from the sum of all reconstructed objects forming the ROE, its $\Delta E$, computed as the difference between the total energy of all ROE objects and half the center-of-mass energy; the numbers of muon, pion and electron candidates in the ROE, which are identified according to the highest $\mathcal{P}_{\mu,\pi,e}$ value; the product of the ROE charge (which can be anything)  and $\tau$ signal charge (which is equal to $\pm1$); the thrust and the following quantities related to the ROE thrust axis, where the ROE thrust axis is computed in a manner similar to the event thrust axis, but using only the ROE detector objects: the cosine of the angle between the ROE thrust axis and the signal thrust axis, and the cosine of the angle between the ROE thrust axis and the $z$ axis.
The third category of input variables comprises the thrust, the numbers of tracks and photons in the event, the total photon energy in the c.m.\ frame, and variables related to the missing momentum of the event: its transverse component in the c.m.\ frame, its polar angle in the c.m.\ frame, the cosine of the angle between the missing momentum and each of the three muons, and the invariant missing mass squared.

The BDT is trained on 176000 simulated signal events and 3400 simulated background $\epem\to\tautau$ and $\epem\to\qqbar$ events in the $\pm 20\,\delta$ region, corresponding to a data set equivalent to an integrated luminosity of 4\invab.
Given that the background retention is already low before the BDT training, a cross-validation (\emph{k-folding}) algorithm is used to reduce the impact of statistical fluctuations on the training results~\cite{10.1145/307400.307439}.
The BDT parameters are optimized using the \texttt{Optuna} library~\cite{akiba_optuna_2019}, which minimizes the logarithmic loss function.
The requirement on the BDT output ($p^{\rm BDT}$) is chosen to maximize the Punzi figure-of-merit~\cite{Punzi:2003bu}  defined as $\varepsilon_{3\mu}/(\alpha /2 + \sqrt{B})$, where $\varepsilon_{3\mu}$
is the signal efficiency, $B$ the number of background events in the signal region and $\alpha=3$ (i.e.\ we optimise the search for a three-standard-deviation signal). The size of the elliptical signal region is also optimized, and the initial width of 5\,$\delta$ is found to be the best one, together with a requirement on the $p^{\rm BDT}$ to be larger than 0.9.
Finally, a data-driven requirement is introduced only to keep events in which the sum of the charged-particles charges is zero. The total signal efficiency evaluated on an independent simulation sample is $(20.42\pm0.06)\%$, where the uncertainty is due to the limited size of the sample.
These values include corrections that account for the differences between data and simulation in particle identification efficiencies.
After all background rejection steps, 88\% of the selected signal candidates in the simulation have one track in the ROE, while 12\% contain three tracks.

\subsection{Simulation validation and expected background events}
\begin{figure}[!ht]
    \centering    
    \includegraphics[width=0.6\textwidth,page=3]{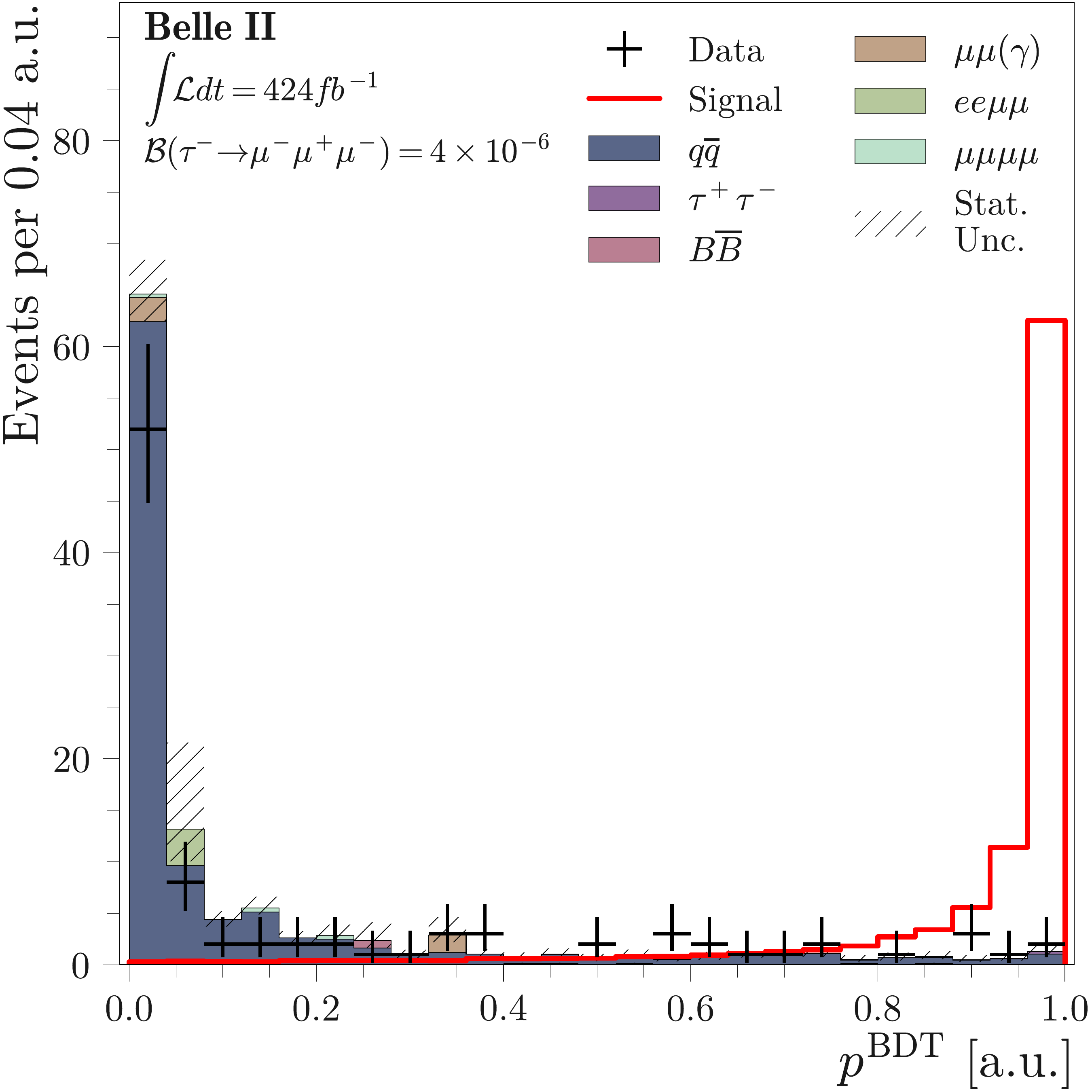}
    \caption{Comparison between data (black points with error bars) and simulation for the BDT output distribution for events in the sideband region after the preselection. The various simulated background processes are shown as a stack of color-filled histograms, with statistical uncertainties displayed as hatched areas, while the signal is shown as a red histogram with an arbitrary scale. } 
    \label{fig:BDT_SB}
\end{figure}

The agreement between data and simulation for the BDT output is checked using the SB region as shown in Fig.~\ref{fig:BDT_SB}.
After the BDT selection,  the number of background events expected in the SB region from the simulation is $2.0^{+0.7}_{-0.5}$, while in data we observe $3$.

\begin{figure}[!ht]
    \centering    
        \includegraphics[width=0.49\textwidth,page=4]{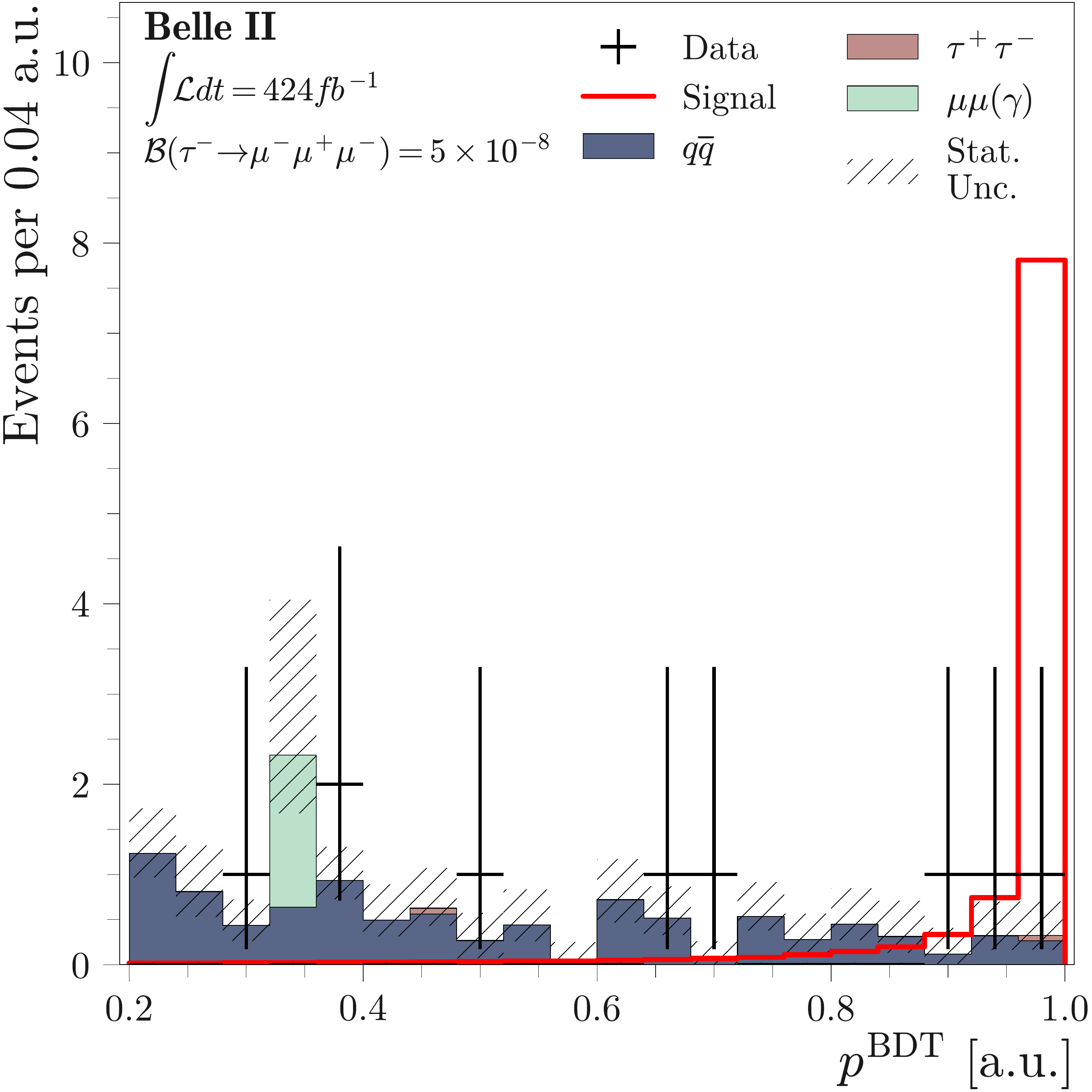}
      \includegraphics[width=0.49\textwidth,page=5]{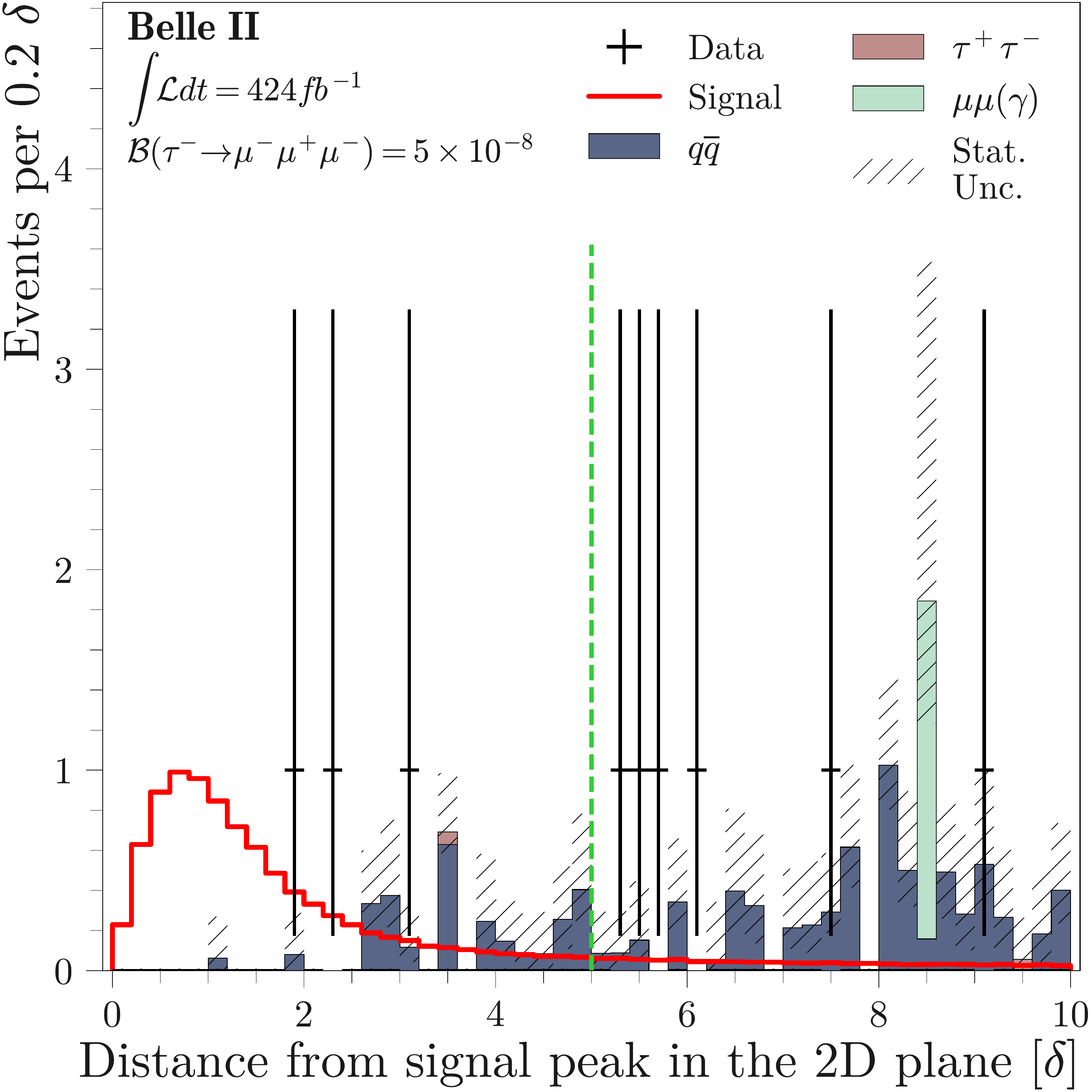}
    \caption{Comparison between data (black points with error bars) and simulation for the BDT output (left) and distance to the signal peak in the (\Mmu, \dE) plane normalised by their respective resolution $\delta$ (right).  The various simulated background processes are shown as a stack of color-filled histograms, with statistical uncertainties displayed as hatched areas, while the signal is shown as a red histogram with an arbitrary scale. The green dashed line in the right-hand plot corresponds to the boundary of the SR. }
    \label{fig:ABCD}
\end{figure}

The number of expected events in the signal region after the BDT selection is obtained from a data-driven method using three other regions of the plane consisting of the BDT output and the distance to the signal peak in the two-dimensional plane: region $A$ is outside the SR with $0.2<p^{\rm BDT}<0.85$; region $B$ is inside the SR with $0.2<p^{\rm BDT}<0.85$; and region $C$ is outside the SR with $p^{\rm BDT}>0.9$. The distributions of events in data and simulation for those two variables are shown in Fig.~\ref{fig:ABCD}.
Events in region $B$ correspond to a subset of the originally hidden $\pm5\,\delta$ SR. However, due to the BDT requirement, the signal efficiency in this region is 50 times lower than in the search region, and the potential signal is negligible compared with the background.
The number of expected events in the  $\pm5\,\delta$ SR  region  is given by 
\begin{equation}
    N_{exp} = N_{C}\times R_{B/A},
\end{equation}
where $R_{B/A} = N_{B}/N_{A}$ is the transfer factor between the region $A$ and $B$.
The measured event yields are 
$N_{A}=4$
, $N_{B}=2$
, and $N_{C}=1$
, resulting in $N_{exp}=0.7_{-0.5}^{+0.6}$,
obtained from pseudoexperiments assuming Poisson distributions with means corresponding to the yields in regions $A$, $B$ and $C$.

%% file: 04-sys.tex
\section{Systematic uncertainties}
\label{sec:systematics}
One category of systematic uncertainties arises from differences between experimental data and simulation due to possible mismodeling in the generation and reconstruction of the simulated samples, and affects the signal efficiency uncertainty.

We take into account the systematic uncertainty associated with the corrections to the simulated muon-identification efficiencies, derived from auxiliary measurements in data using $\jpsi\to \mup\mun$, $\epem\to\mup\mun\gamma$, and $\epem\to\epem\mup\mun$ events. These corrections are obtained as functions of momentum, polar angle and charge, and applied to events reconstructed from simulation. The systematic uncertainty is obtained by varying the corrections within their statistical and systematic uncertainties and estimating the impact of these variations on the selection efficiency. Adding the statistical and systematic variations in quadrature, the result is a relative uncertainty in the signal efficiency of $^{+2.4}_{-2.1}$\%.

The difference between data and simulation in track-reconstruction efficiency is measured  in \(\epem\to\taup\taum\) events, selecting $\taum\to \en\neueb\neut$  and  $\taum\to\pim\pip\pim\neut$ decays.
A discrepancy of 0.24\% per track is observed, resulting in a systematic uncertainty of $1.0\%$.

The agreement between data and simulation for the trigger efficiency is evaluated using the $\taum\to\pim\pip\pim\neut$ control sample.
In data, the trigger efficiency is computed using independent trigger selections: the efficiency of the ECL-based trigger selection is obtained using events triggered by the CDC, while the efficiency of the CDC-based trigger selection is evaluated using events passing the ECL trigger requirements. The agreement between data and simulation efficiencies is 0.5\% for the ECL trigger selection and is 4.3\% for the CDC trigger selection. Given that the efficiency of trigger selections based on the ECL only is 88\%, the weighted average of the discrepancies is computed to be 0.7\% and this is used as the systematic uncertainty of the trigger efficiency. 
The possible bias coming from the use of independent trigger selections to evaluate the trigger efficiency is tested on simulated events. A difference of 0.5\% is found with respect to the absolute efficiency,  and added in quadrature to the previous uncertainty.

The $\taum\to\pim\pip\pim\neut$ control sample is also used to obtain a systematic uncertainty on the BDT selection. The same BDT that was trained for the signal decay is applied to these events and a requirement is chosen on the output in order to have the same efficiency on $\taum\to\pim\pip\pim\neut$ events as on the signal decay. The difference between the efficiency in
data and simulation for this BDT requirement, $1.5\%$, is considered as a systematic uncertainty.

In order to take into account possible mismodeling of the ISR, FSR, and resolution  of the signal extraction variables, we change the signal region definition by $\pm 1 \delta$. This results in a $^{+2.9}_{-3.9}\%$ variation of the signal efficiency, that we take as a systematic uncertainty.

The number of expected background events is affected by an uncertainty that originates from imperfections in the magnetic field description used in the event reconstruction and material mismodeling. 
To correct for those effects, the momenta of charged particles are scaled with a factor of 0.99987,  evaluated by measuring the mass-peak position of high-yield samples of $\Dz\to\Km\pip$ decays reconstructed in data and comparing it to the world average value of the $D^0$ mass~\cite{ParticleDataGroup:2022pth}. The corresponding systematic uncertainty is obtained by varying the correction factor according to its uncertainty ($^{+3.8}_{-5.7}\times 10^{-4}$), which leads to different data yields in the sideband region and, thus, different numbers of expected events $N_{\mathrm{exp}}$. The resulting systematic uncertainty, taken as the difference from the nominal value, is $16$\%, and is mainly due to the small number of events.

The systematic uncertainty in the integrated luminosity $\mathcal{L}$, measured with samples of Bhabha and diphoton events~\cite{Belle-II:2019usr}, is evaluated from the difference observed between the results from the two methods and amounts to 0.6\% relative uncertainty. 

Finally, we also assign an uncertainty of 0.003 nb on the $\tau$-pair production cross section, as evaluated in Ref.~\cite{Banerjee:2007is}. 

A summary of the systematic uncertainties is given in Table~\ref{tab: systematics}.
\begin{table}[htbp]
    \centering
\caption{Summary of relative systematic uncertainties.}
\begin{tabular}{c|c|cc}
\hline\hline
   &  &  \multicolumn{2}{c}{Uncertainty (\%)} \\
     Quantity  & Source  & Low & High \\
\hline
 \multirow[c]{6}{*}{\(\varepsilon_{\mathrm{3\mu}}\)} & PID &  2.1 & 2.4 \\
   & Tracking  & 1.0 & 1.0 \\
   & Trigger  & 0.9 & 0.9 \\
   & BDT &  1.5 & 1.5 \\
   & Signal region  & 3.9 & 2.9 \\
    &&&\\
    \(N_{exp}\) & Momentum Scale  & 16 & 16 \\
 &&&\\
 \(\mathcal{L}\) &   & 0.6 & 0.6 \\
 &&&\\
 \(\sigma_{\tau\tau}\) & & 0.3 & 0.3 \\
\hline\hline
\end{tabular}
\label{tab: systematics}
\end{table}

In order to perform a validation of the inclusive reconstruction and BDT selection, we carry out an alternative analysis using one-prong tagging reconstruction similar to that done previously by Belle and Babar.
We select events containing exactly four charged particles with zero total charge.
Within the tag side, we categorize the charged particle as either a lepton (leptonic tag) or a hadron (hadronic tag).
The contamination from $\epem \to \epem nh$ events is mitigated using the data-driven requirements $0.4<\theta_{{\rm{miss}}}^*<2.7$\rad for leptonic and $0.3<\theta_{{\rm{miss}}}^*<2.8$\rad for hadronic tags, along with a requirement at most one signal track points to the endcaps. 
These requirements also suppress the $\epem \to \ell^+\ell^-\ell^+\ell^-$ and $\epem \to \epem h^+h^-$ background components, which are further rejected by requiring the difference between the tag tau energy and half the center-of-mass energy to be less than $-0.8$\gev,  and $0.90 < T< 0.97$. 
The \eeqqbar background is suppressed by requiring the visible c.m.\ energy of all reconstructed particles in the event $\visE$ to be smaller than 10.2\gev and the missing momentum $p_{\rm{miss}}^*$ to be larger than 0.4 \gevc.
To optimize the muon identification selection, charged particles are separated according to their momentum into three categories corresponding to tracks not reaching, partially crossing, or fully crossing the KLM.
Selection criteria are optimized independently for the three categories using the Punzi figure-of-merit.
The signal efficiency is 14.9\%, which is nearly twice that achieved at the first generation \B-factory experiments. This increase is due to a more optimal selection, which includes low momentum muons on the signal side and $\tau^-\to\mu^-\nu\bar\nu$ decays in the tag side, as well as improved muon identification.
The number of background events expected from simulation is 0.43. 
The inclusive-tagging selection has a 37\% higher efficiency than that of the one-prong tagging method for a similar level of background and is thus used as the final result.

%% file: 05-UL.tex
\section{Result}
\label{sec:results}
\begin{figure}
    \centering
   \includegraphics[page=6,width=0.6\columnwidth]{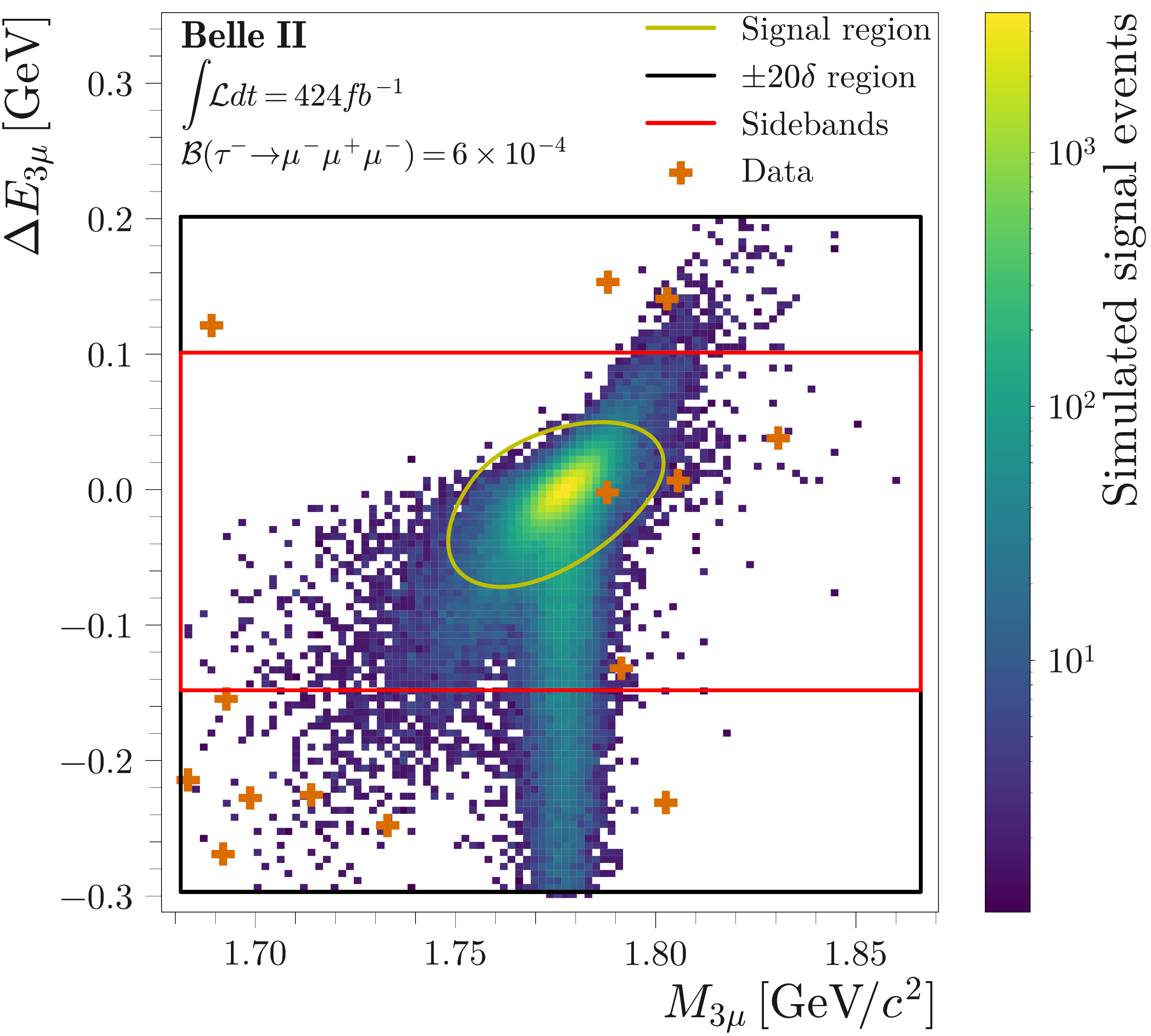}
    \caption{Scatter plot of selected events in the (\Mmu, \dE) plane for data (orange crosses) and simulated signal (color-filled area). The SB region is shown as the red rectangle.  The yellow ellipse represents the signal region. Events outside the black rectangle, representing the $\pm20\delta$ region, are discarded.}
    \label{fig: Yields extraction sideband method}
\end{figure}

\begin{figure}
    \centering
    \includegraphics[width=0.49\textwidth,page=7]{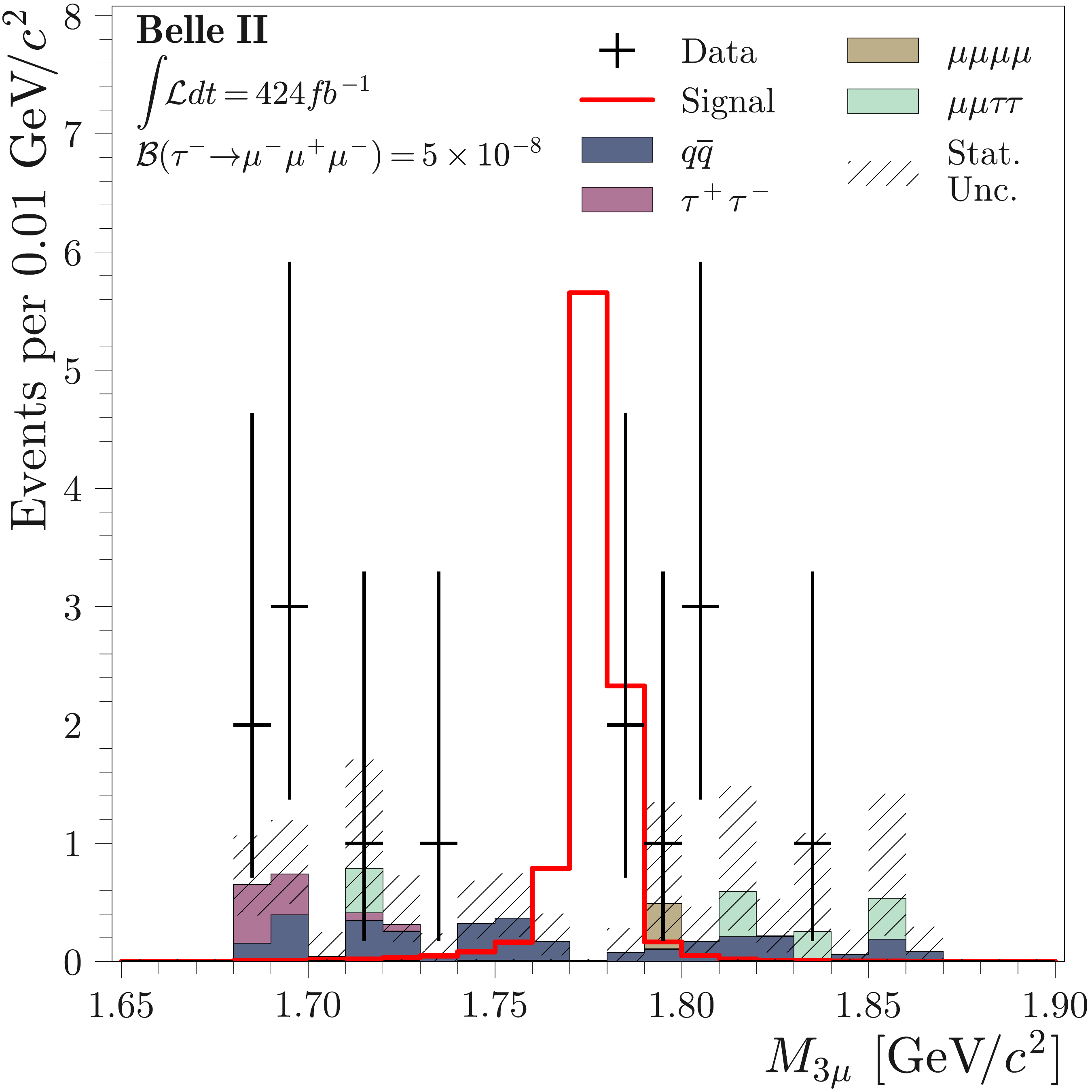}
    \includegraphics[width=0.49\textwidth,page=8]{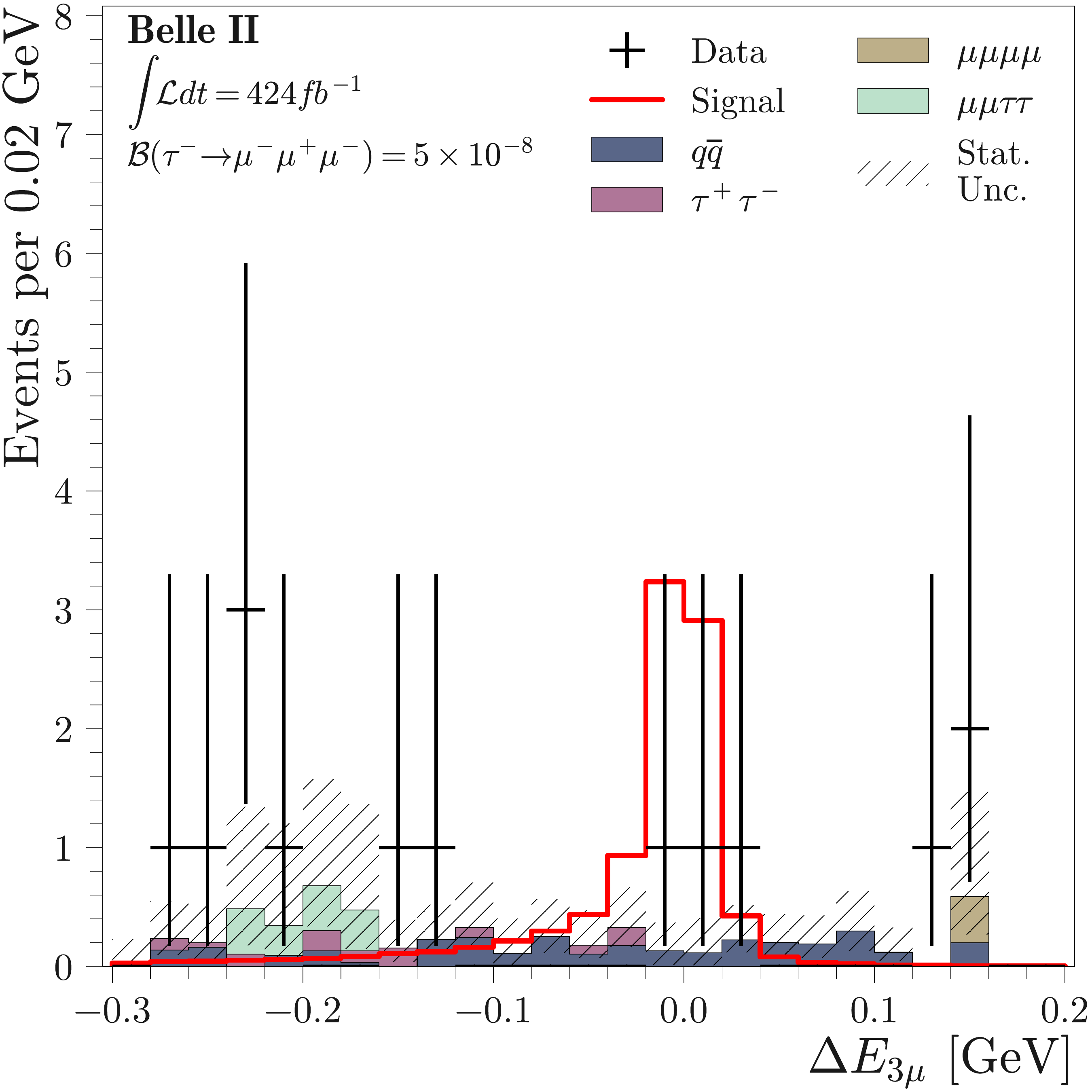}
    \caption{Comparison between data (black points with error bars) and simulation for the \Mmu and \dE distributions. Events are those populating the entire black rectangle shown in Fig.~\ref{fig: Yields extraction sideband method}. The various simulated background processes are shown as a stack of color-filled histograms, while the signal is shown as a red histogram with an arbitrary scale. The statistical uncertainties are displayed as hatched areas.}
    \label{fig: Inclusive DataMC comparison for best M and DeltaE}
\end{figure}

The distribution of events in the (\Mmu, \dE) plane is shown in Fig.~\ref{fig: Yields extraction sideband method} for data and simulated signal events. We observe one event in the signal region, $N_{obs}=1$.
Using a signal efficiency of $\varepsilon_{\mathrm{3\mu}} = (20.42\pm0.06^{+1.02}_{-0.84})\%$, where the first uncertainty is statistical and the second are systematic, and a number of expected background events $N_{exp}=0.7_{-0.5-0.1}^{+0.6+0.1}$ in the formula given in Eq.~\ref{eq:result},
\begin{equation}\label{eq:result}
 \mathcal{B}(\taumu) = \frac{N_{obs}-N_{exp}}{\mathcal{L} \times 2\sigma_{\tau\tau} \times \varepsilon_{\mathrm{3\mu}}},
\end{equation}
we compute a branching fraction of $\mathcal{B}(\taumu)= (2.1^{+5.1}_{-2.4}\pm0.4)\times10^{-9}$, where the first uncertainty is statistical and the second is systematic.
The other inputs to Eq.~\ref{eq:result} are the integrated luminosity $\mathcal{L}=424\pm 3~\invfb$ of the analyzed data sample; the $\tau$-pair production cross section $\sigma_{\tau\tau}=0.919\pm 0.003~\nb$, where a weighted average of the cross sections at the different data taking energies is used.
The distributions of all selected events in the $\pm 20\,\delta$ region for data and simulation for the \Mmu and \dE variables are shown in Fig.~\ref{fig: Inclusive DataMC comparison for best M and DeltaE}.

\begin{figure}[htb]
    \centering
   \includegraphics[width=0.60\columnwidth]{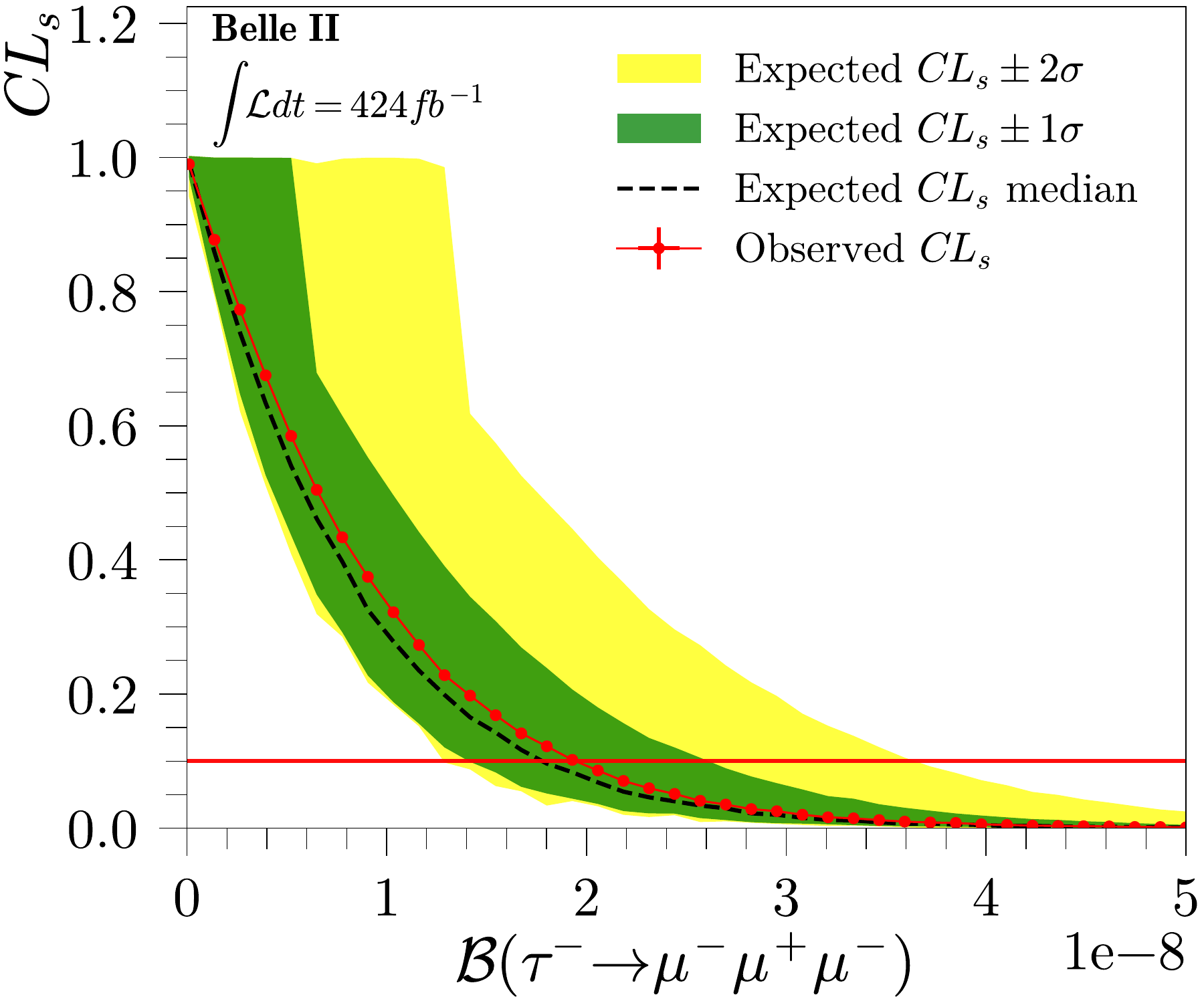}
    \caption{Observed (solid red curve with dots) and expected (dashed black curve) $\mathrm{CL}_s$ as a function of the assumed branching fractions of \taumu. The red line corresponds to the 90\% C.L upper limit.}
    \label{fig:CLs}
\end{figure}

As no signal is found, we compute a 90\% confidence level (C.L.) upper limit on the \taumu branching fraction.
We estimate the upper limit using the modified frequentist  $\mathrm{CL}_s$~\cite{Junk:1999kv, Read:2002hq} method implemented in the RooStat framework.
We generate $5\times 10^4$ pseudo-experiments at 40 points distributed uniformly in the branching fraction range $(0 - 5) \times 10^{-8}$. The total statistical and systematic uncertainties affecting each experimental input, discussed in Sect.~\ref{sec:systematics}, are combined in quadrature. Figure \ref{fig:CLs} shows the $\mathrm{CL}_s$ curves computed as a function of the upper limit on the branching fractions for the inclusive tagging analysis. The dashed black line shows the expected $\mathrm{CL}_s$ and the green and yellow bands give the $\pm1\sigma$ and $\pm2\sigma$ contours, respectively. 
The expected limit assuming  an observed number of events equal to 0.7, as expected from background estimation, is $1.8\times 10^{-8}$ at 90\% C.L, while 
the observed limit on the branching fraction of \taumu is $1.9\times 10^{-8}$ at 90\% C.L.

The limit is also evaluated using a Bayesian approach \cite{CALDWELL20092197} yielding the same result.
Using the alternative one-prong tagging analysis, 
zero events are observed in the signal region. This corresponds to a branching fraction of $\mathcal{B}(\taumu)= (-3.7^{+1.9}_{-0.2}\pm 0.1)\times 10^{-9}$, where the first uncertainty is statistical and the second is systematic and takes into account the correction to the muon identification, the trigger and tracking efficiencies.
The small statistical uncertainty is coming from the fact that the number of expected background events is derived from simulation. The corresponding limit is $2.0\times 10^{-8}$ at 90\% C.L.

%% file: 06-summ.tex
\section{Summary}
\label{sec:summ}
We present a search for the LFV decay \taumu using a 424\invfb sample of electron-positron collision data collected by the Belle II experiment.
Using a novel inclusive-tagging reconstruction followed by a BDT-based selection, the efficiency is higher by a factor of 2.5 than the efficiency in the latest Belle analysis~\cite{Hayasaka:2010np} and 37\% higher than a one-prong tagging reconstruction performed on the same Belle II dataset, for an expected number of background events compatible with zero.
We observe one event in the signal region, which corresponds to a branching fraction of $\mathcal{B}(\taumu)= (2.1^{+5.1}_{-2.4}\pm0.4)\times10^{-9}$, where the first uncertainty is statistical and the second one systematic.
The observed (expected) limit at 90\% C.L. computed in a frequentist approach is 1.9 (1.8) $\times10^{-8}$, which is more restrictive than the previous lowest limit.

%% file: acknowledgements-b2.tex
This work, based on data collected using the Belle II detector, which was built and commissioned prior to March 2019,
was supported by
Higher Education and Science Committee of the Republic of Armenia Grant No.~23LCG-1C011;
Australian Research Council and Research Grants
No.~DP200101792, 
No.~DP210101900, 
No.~DP210102831, 
No.~DE220100462, 
No.~LE210100098, 
and
No.~LE230100085; 
Austrian Federal Ministry of Education, Science and Research,
Austrian Science Fund
No.~P~34529,
No.~J~4731,
No.~J~4625,
and
No.~M~3153,
and
Horizon 2020 ERC Starting Grant No.~947006 ``InterLeptons'';
Natural Sciences and Engineering Research Council of Canada, Compute Canada and CANARIE;
National Key R\&D Program of China under Contract No.~2022YFA1601903,
National Natural Science Foundation of China and Research Grants
No.~11575017,
No.~11761141009,
No.~11705209,
No.~11975076,
No.~12135005,
No.~12150004,
No.~12161141008,
and
No.~12175041,
and Shandong Provincial Natural Science Foundation Project~ZR2022JQ02;
the Czech Science Foundation Grant No.~22-18469S;
European Research Council, Seventh Framework PIEF-GA-2013-622527,
Horizon 2020 ERC-Advanced Grants No.~267104 and No.~884719,
Horizon 2020 ERC-Consolidator Grant No.~819127,
Horizon 2020 Marie Sklodowska-Curie Grant Agreement No.~700525 ``NIOBE''
and
No.~101026516,
and
Horizon 2020 Marie Sklodowska-Curie RISE project JENNIFER2 Grant Agreement No.~822070 (European grants);
L'Institut National de Physique Nucl\'{e}aire et de Physique des Particules (IN2P3) du CNRS
and
L'Agence Nationale de la Recherche (ANR) under grant ANR-21-CE31-0009 (France);
BMBF, DFG, HGF, MPG, and AvH Foundation (Germany);
Department of Atomic Energy under Project Identification No.~RTI 4002,
Department of Science and Technology,
and
UPES SEED funding programs
No.~UPES/R\&D-SEED-INFRA/17052023/01 and
No.~UPES/R\&D-SOE/20062022/06 (India);
Israel Science Foundation Grant No.~2476/17,
U.S.-Israel Binational Science Foundation Grant No.~2016113, and
Israel Ministry of Science Grant No.~3-16543;
Istituto Nazionale di Fisica Nucleare and the Research Grants BELLE2;
Japan Society for the Promotion of Science, Grant-in-Aid for Scientific Research Grants
No.~16H03968,
No.~16H03993,
No.~16H06492,
No.~16K05323,
No.~17H01133,
No.~17H05405,
No.~18K03621,
No.~18H03710,
No.~18H05226,
No.~19H00682, 
No.~20H05850,
No.~20H05858,
No.~22H00144,
No.~22K14056,
No.~22K21347,
No.~23H05433,
No.~26220706,
and
No.~26400255,
and
the Ministry of Education, Culture, Sports, Science, and Technology (MEXT) of Japan;  
National Research Foundation (NRF) of Korea Grants
No.~2016R1\-D1A1B\-02012900,
No.~2018R1\-A2B\-3003643,
No.~2018R1\-A6A1A\-06024970,
No.~2019R1\-I1A3A\-01058933,
No.~2021R1\-A6A1A\-03043957,
No.~2021R1\-F1A\-1060423,
No.~2021R1\-F1A\-1064008,
No.~2022R1\-A2C\-1003993,
and
No.~RS-2022-00197659,
Radiation Science Research Institute,
Foreign Large-Size Research Facility Application Supporting project,
the Global Science Experimental Data Hub Center of the Korea Institute of Science and Technology Information
and
KREONET/GLORIAD;
Universiti Malaya RU grant, Akademi Sains Malaysia, and Ministry of Education Malaysia;
Frontiers of Science Program Contracts
No.~FOINS-296,
No.~CB-221329,
No.~CB-236394,
No.~CB-254409,
and
No.~CB-180023, and SEP-CINVESTAV Research Grant No.~237 (Mexico);
the Polish Ministry of Science and Higher Education and the National Science Center;
the Ministry of Science and Higher Education of the Russian Federation
and
the HSE University Basic Research Program, Moscow;
University of Tabuk Research Grants
No.~S-0256-1438 and No.~S-0280-1439 (Saudi Arabia);
Slovenian Research Agency and Research Grants
No.~J1-9124
and
No.~P1-0135;
Agencia Estatal de Investigacion, Spain
Grant No.~RYC2020-029875-I
and
Generalitat Valenciana, Spain
Grant No.~CIDEGENT/2018/020;
National Science and Technology Council,
and
Ministry of Education (Taiwan);
Thailand Center of Excellence in Physics;
TUBITAK ULAKBIM (Turkey);
National Research Foundation of Ukraine, Project No.~2020.02/0257,
and
Ministry of Education and Science of Ukraine;
the U.S. National Science Foundation and Research Grants
No.~PHY-1913789 
and
No.~PHY-2111604, 
and the U.S. Department of Energy and Research Awards
No.~DE-AC06-76RLO1830, 
No.~DE-SC0007983, 
No.~DE-SC0009824, 
No.~DE-SC0009973, 
No.~DE-SC0010007, 
No.~DE-SC0010073, 
No.~DE-SC0010118, 
No.~DE-SC0010504, 
No.~DE-SC0011784, 
No.~DE-SC0012704, 
No.~DE-SC0019230, 
No.~DE-SC0021274, 
No.~DE-SC0021616, 
No.~DE-SC0022350, 
No.~DE-SC0023470; 
and
the Vietnam Academy of Science and Technology (VAST) under Grants
No.~NVCC.05.12/22-23
and
No.~DL0000.02/24-25.

These acknowledgements are not to be interpreted as an endorsement of any statement made
by any of our institutes, funding agencies, governments, or their representatives.

We thank the SuperKEKB team for delivering high-luminosity collisions;
the KEK cryogenics group for the efficient operation of the detector solenoid magnet;
the KEK Computer Research Center for on-site computing support; the NII for SINET6 network support;
and the raw-data centers hosted by BNL, DESY, GridKa, IN2P3, INFN, 
and the University of Victoria.